\documentclass[prl,twocolumn]{revtex4}
\usepackage{graphicx}
\usepackage{amssymb,amsmath}
\usepackage{enumerate}

\newcommand{\be}{\begin{eqnarray}}
\newcommand{\ee}{\end{eqnarray}}

\begin{document}

\title{Hamiltonian distributed chaos in the Asian-Australian Monsoons and in the ENSO}

\author{A. Bershadskii}

\affiliation{
ICAR, P.O. Box 31155, Jerusalem 91000, Israel
}

\begin{abstract}
Two subsystems of the Asian Monsoon: the Indian Summer Monsoon and the Western North Pacific Monsoon, have been analysed using their daily indices ISMI and WNPMI. It is shown that on the intraseasonal time scales the ISMI and WNPMI are dominated by the Hamiltonian distributed chaos  with the stretched exponential spectrum: $E(f) \propto \exp-(f/f_0)^{\beta}$ and analytical values of the parameter $\beta =3/4$ and $\beta =1/2$ correspondingly. The relevant daily indices Ni\~no 3 and Ni\~no 4 (with $\beta =1/2$) of the El Ni\~no-Southern Oscillation (ENSO) and the Australian Monsoon (AUSM index with $\beta = 1/2$) have been also discussed in this context.

\end{abstract}

\maketitle

\section{The Asian Monsoon}

\begin{figure}
\begin{center}
\includegraphics[width=8cm \vspace{-0cm}]{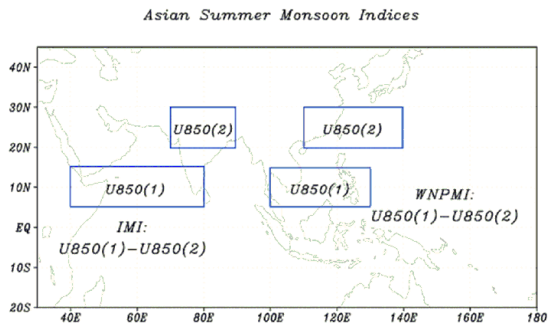}\vspace{-6.5cm}
\caption{\label{fig1} Indian Summer Monsoon Index (left) and  Western North Pacific Monsoon Index (right) \cite{mon}. } 
\end{center}
\end{figure}
\begin{figure}
\begin{center}
\includegraphics[width=8cm \vspace{-0.3cm}]{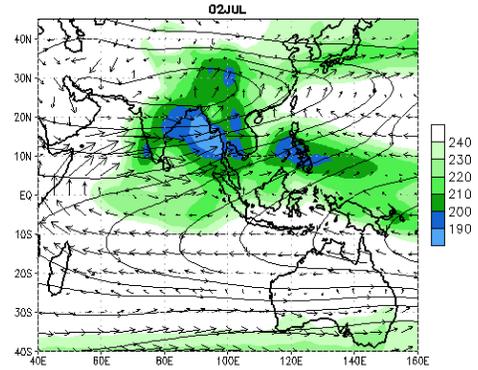}\vspace{-5cm}
\caption{\label{fig2} Asian-Australian Monsoons on 2 Jul.: ORL, 200-hPa Streamlines and 850-hPa Wind Clim (1979-1995) \cite{global}.} 
\end{center}
\end{figure}
  Two dominant regions of energy supply (by convective heat) can be determined for the Asian monsoon: the Philippine Sea and Bay of Bengal (see for a comprehensive review Ref. \cite{wwl}). 
Weak correlation between these rather different energy sources resulted in appearance of two different in their properties monsoon indices: Western North Pacific Monsoon Index (WNPMI) and Indian Summer Monsoon Index (ISMI or IMI) \cite{wwl},\cite{wf}. Together these indices describe one of the most powerful oscillating pattern of the earth's climate. Definition of ISMI and WNPMI can be understood from the figure 1 \cite{wwl},\cite{mon}. The regions where the zonal winds are used for computation of the monsoon circulation indices are denoted in the Fig. 1 by the boxes (see also figure 2 \cite{global}). 
According to the Refs. \cite{wwl},\cite{wf} the difference of the 850-hPa zonal winds is used for the definition of the
$$
\scriptstyle ISMI= U850(40E-80E,5N-15N)-U850(70E-90E,20N-30N)
$$
and the difference of 850-hPa westerlies is used for the definition of the
$$
\scriptstyle WNPMI = U850(100E-130E,5N-15N)-U850(110E-140E,20N-30N) 
$$ 
The ISMI represents the rainfall anomalies over a region including the India, Bay of Bengal and
the eastern Arabian Sea. The WNPMI represents also the low-level vorticity generated by response of the Rossby waves to convective heat source located at the Philippine Sea (cf. the Ref. \cite{b}). 

  Correlation between these indices is very weak that corresponds well to the difference in their origins. Dissimilarity between the ISM and WNPM geographic setting in respect of the
ocean-continent distribution results in the considerable differences in their variability for all time scales. 

  Despite this, convective activity related to the Asian Monsoon results in
a global pattern expanding in both hemispheres for intraseasonal as well as for inter-annual time scales \cite{lin},\cite{ding}. Even Arctic ice patterns can be driven by the Asian Summer Monsoon via North Atlantic \cite{gw}.

\section{Hamiltonian distributed chaos}

\begin{figure}
\begin{center}
\includegraphics[width=8cm \vspace{-0.6cm}]{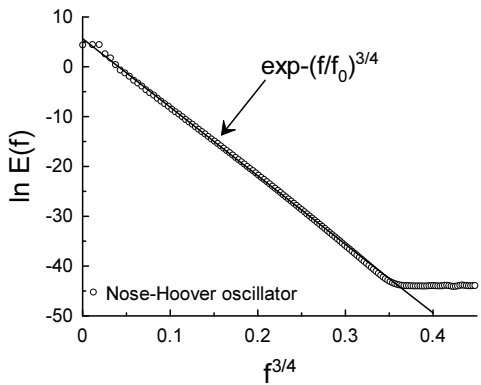}\vspace{-4.2cm}
\caption{\label{fig3} Power spectrum of the $x$ fluctuations for the Nose-Hoover oscillator. } 
\end{center}
\end{figure}
  
\begin{figure}
\begin{center}
\includegraphics[width=8cm \vspace{-0.25cm}]{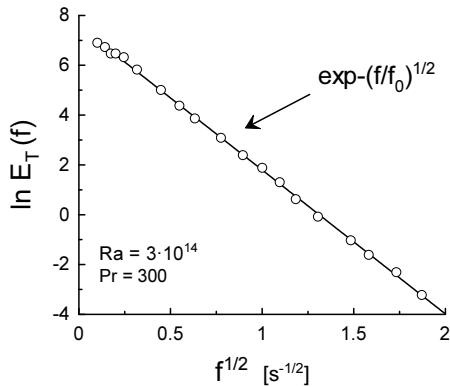}\vspace{-4.5cm}
\caption{\label{fig4} Temperature power spectrum in a thermal convective cell (the data were taken from Ref. \cite{as}). } 
\end{center}
\end{figure}

   Chaotic dynamical systems typically exhibit exponential power spectra (see, for instance \cite{fm}-\cite{sig}):
$$
 E(f) \propto \exp-(f/f_c) \eqno{(1)}
$$ 
  A more complex situation takes place for analytical Hamiltonian dynamical systems. Namely, a weighted superposition of the exponentials produces a stretched exponential high-frequency spectral tails
$$
E(f ) \propto \int_0^{\infty} P(f_c) \exp -(f/f_c)~ df_c  \propto \exp-(f/f_0)^{\beta}  \eqno{(2)}
$$
with $\beta =3/4$ or $\beta =1/2$ \cite{b1}. The value $\beta =1/2$ can also appear due to adiabatic invariance of the action \cite{suz} at spontaneous breaking of the time translational symmetry for the Hamiltonian dynamical systems \cite{b2}. The spectrum Eq. (2) represents the Hamiltonian distributed chaos.  

  Figure 3 shows, as an example, power spectrum for the $x(t)$ variable of the Nose-Hoover oscillator (the Sprott A system \cite{spot}). This oscillator can be considered as a harmonic oscillator contacting with a thermal bath and can be described by the system of equations
$$
\left\{ \begin{array}{l} \dot{x} = y \\[0.1cm] \dot{y} = -x + yz \\[0.1cm] \dot{z} = 1 - y^2 \end{array} \right.  \eqno{(3)}
$$
The thermostat is represented by the term - $yz$. The dot over a variable denotes a time derivative. The system is a Hamiltonian one \cite{spot}.

 The data for computation of the spectrum shown in the Fig. 3 were taken from the site \cite{gen}. The computation was performed using the maximum entropy method, providing an optimal resolution for chaotic time series \cite{oh}. The straight line in the Fig. 3 is drawn to indicate correspondence (in the appropriately chosen scales) to the Eq. (2) with $\beta = 3/4$. \\
 
   Another example of the spectrum Eq. (2) (now with $\beta =1/2$) is shown in figure 4. This figure shows a temperature power spectrum.  The temperature was measured at the center of an upright cylinder cell with  strong thermal convection at very large Rayleigh number $Ra=3\cdot 10^{14}$ \cite{as} (see Ref. \cite{tg},\cite{gl} describing the  low-order Hamiltonian models of the Rayleigh-Benard convection).

 \begin{figure}
\begin{center}
\includegraphics[width=8cm \vspace{-0.45cm}]{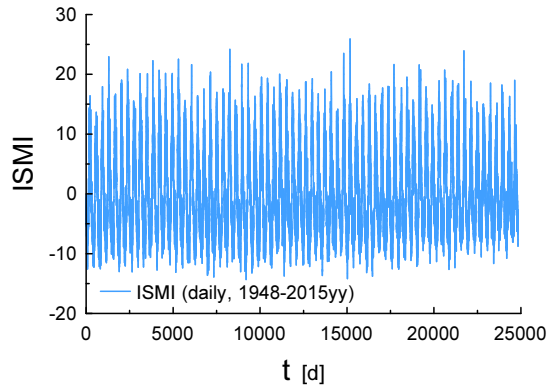}\vspace{-4.5cm}
\caption{\label{fig5} Daily ISM index for the period 1948-2015yy \cite{mon} } 
\end{center}
\end{figure}  
   
\section{Chaotic ISM and WNPM indices }

Most of the theoretical models constructed in the geophysical fluid dynamics are Hamiltonian \cite{b},\cite{she}-\cite{gl2} and it has been already established that the Hamiltonian distributed chaos determines the AAO, AO, NAO and PNA daily indices on the intraseasonal scales \cite{b2},\cite{b3}. Therefore, one can expect the same situation for the Asian Monsoon daily indices as well. 

  In the time series the intraseasonal time scales can be separated by removing of the components of interannual variability and annual cycle. Subtraction of a 120-day moving average is considered as an appropriate tool for removing the low frequency variations in such cases (see, for instance, Ref. \cite{ven2} and references therein). We have used the 121-day moving average for this purpose: 
$$ 
y(i)=\frac{\sum_{i-k}^{i+k} x(i)}{(2k+1)}  \eqno{(4)}
$$

Figure 5 shows the daily ISM index for the period 1948-2015yy (the data were taken from the Ref. \cite{mon}). Figure 6 shows power spectrum for the intraseasonal (i.e. after the subtraction) daily ISM index for the period 1948-2015yy (the data for the computations were taken from the site \cite{mon}). The straight line is drawn to indicate correspondence to the Eq. (2) with $\beta =3/4$. The characteristic time scale obtained from the best fit is $T_0=1/f_0 \simeq 32d$. Figure 7 shows power spectrum corresponding to the intraseasonal daily WNPM index for the period 1948-2015yy (the data for the computations were taken from the site \cite{mon}). The straight line is drawn to indicate correspondence to the Eq. (2) now with $\beta =1/2$. The characteristic time scale obtained from the best fit is $T_0=1/f_0 \simeq 171$d.   \\

  One can see that the both analytical types of the Hamiltonian distributed chaos are present in the Asian Monsoon. The difference between its two components, mentioned in the first Section, is reflected in the difference between the two values of the $\beta$ parameter (cf. Figs. 6 and 7).

\section{Australian Monsoon}

 The Australian Monsoon is a natural equatorial counterpart for the Asian (mainly WNPM) Monsoon. Indeed, flow across the Equator of the dry air from the continent, where a winter (for corresponding hemisphere) takes place, toward the hemisphere with summer conditions delivers moisture, that was taken on the way from the worm oceans. This moisture feeds the monsoon rains at the hemisphere with the summer conditions. The reverse of the winds direction each half a year results in the alternation between the monsoon rainfalls in the Northern and the Southern hemispheres.  \\
 
   The rainfall and wind records at the most northerly of the Australian capital cities - Darwin ($12^o$S, $130^o$E, see figure 8) were used to describe the Australian Monsoon in majority of early studies (see, for instance, the Refs. \cite{kim},\cite{sup} and references therein). However, as one can see from figure 9 a broad scale wind circulation index properly corresponding to the Australian Monsoon should be based at a different location. The authors of the recent paper Ref. \cite{ka} (see also an earlier paper Ref. \cite{wang2}) suggested such index - AUSMI, based on 850 hPa zonal wind anomalies averaged over the region $5^o$S-$15^o$S, $110^o$E-$130^o$E (see the Fig. 8). It should be noted that the Australian monsoon rainfall is located approximately in the area ($7.5^o$S-$17.5^o$S, $120^o$E-$150^o$E), i.e. well including the Darwin location (see also below). Despite this the AUSMI captures well not only the interannual and intraseasonal time scales variability of the Australian monsoonal rainfall but also the relationship of the Australian monsoon with the ENSO \cite{ka}. \\
\begin{figure}
\begin{center}
\includegraphics[width=8cm \vspace{-0.5cm}]{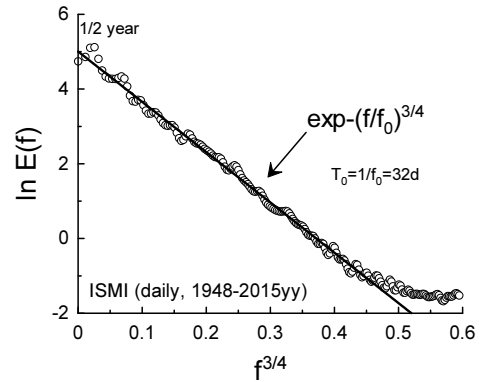}\vspace{-4.5cm}
\caption{\label{fig6} Power spectrum for the intraseasonal daily ISM index. } 
\end{center}
\end{figure}
\begin{figure}
\begin{center}
\includegraphics[width=8cm \vspace{-0.3cm}]{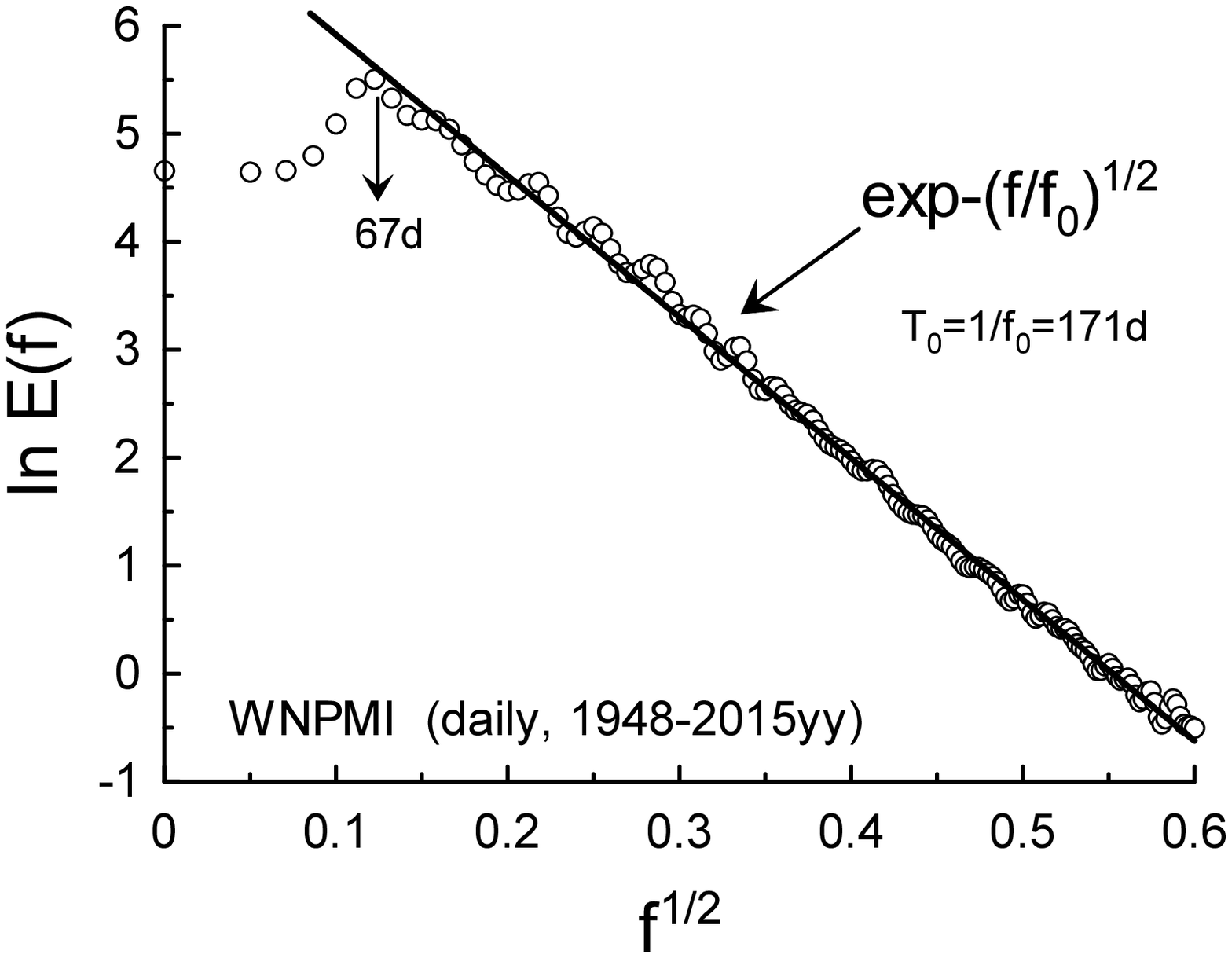}\vspace{-4cm}
\caption{\label{fig6} Power spectrum for the intraseasonal daily WNPM index. } 
\end{center}
\end{figure}
\begin{figure}
\begin{center}
\includegraphics[width=8cm \vspace{-0.6cm}]{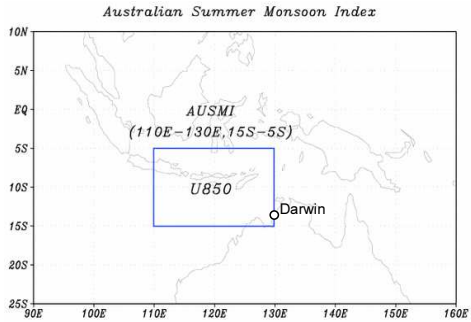}\vspace{-5.5cm}
\caption{\label{fig5} The Australian Monsoon Index - AUSMI (adapted from the site \cite{mon}).} 
\end{center}
\end{figure}  
\begin{figure}
\begin{center}
\includegraphics[width=8cm \vspace{-0.4cm}]{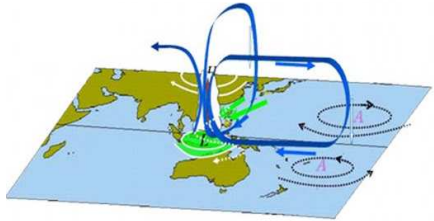}\vspace{-6cm}
\caption{\label{fig5}  A sketch of the broad-scale circulation corresponding to the active Australian Monsoon (adapted from the site \cite{cli}). } 
\end{center}
\end{figure}   
\begin{figure}
\begin{center}
\includegraphics[width=8cm \vspace{-0.1cm}]{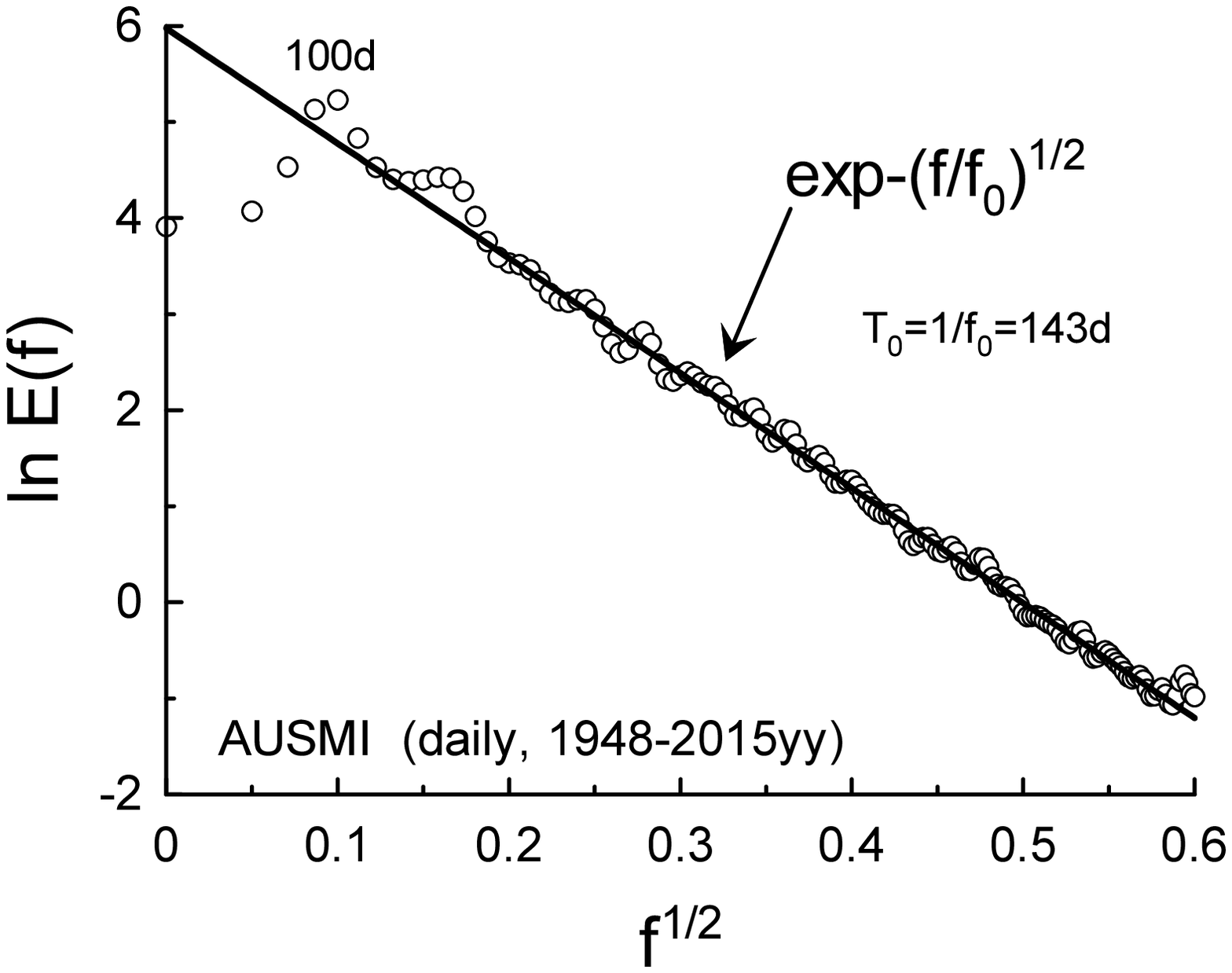}\vspace{-4cm}
\caption{\label{fig5} Power spectrum for the intraseasonal daily AUSM index.} 
\end{center}
\end{figure} 
\begin{figure}
\begin{center}
\includegraphics[width=8cm \vspace{+0.3cm}]{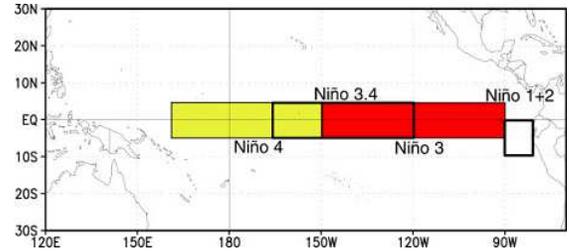}\vspace{-7.3cm}
\caption{\label{fig5} Map of the tropical Pacific regions \cite{reg}.} 
\end{center}
\end{figure}   
  
   Figure 10 shows shows power spectrum corresponding to the intraseasonal daily AUSM index for the period 1948-2015yy (the data for the computations were taken from the site \cite{mon}). The straight line is drawn to indicate correspondence to the Eq. (2) with $\beta =1/2$. The characteristic time scale obtained from the best fit is $T_0=1/f_0 \simeq 143d$. One can see that the AUSMI power spectrum corresponds to that of the WNPMI's type ($\beta=1/2$) rather than that of ISMI's type ($\beta = 3/4$). It could be expected, because the Australian monsoon has geographic setting in respect of the ocean-continent distribution more similar to that corresponding to WNPM rather than that corresponding to ISM.

\section{Relationship with ENSO}

  Naturally relationship between the Asian-Australian Monsoons and El Ni\~no (La Ni\~na)-Southern
Oscillation (ENSO) phenomenon is of great interest (see, for instance, Refs. \cite{wwl},\cite{ka}-\cite{gre} and references therein). The relationships between the ENSO and the two subsystems of the Asian Monsoon are different, according to the difference between these subsystems. While the relationship between the ENSO and the Indian Summer Monsoon is rather uncertain and non-stable the relationship between the ENSO and the NWPM is much more strong and stable. Relationship between the Australian Monsoon and the ENSO is even stronger due to their geographic setting (cf. Figs. 8 and 11, and see below).\\ 
  
    The Ni\~no 3 (5N-5S, 150W-90W region) and the Ni\~no 4 (5N-5S, 160E-150W region) indices - Fig. 11, based on sea surface temperature (SST) anomalies averaged across a given region, can be considered as the relevant ones for comparison with the Asian-Australian Monsoons indices. It should be noted that the Ni\~no 4 region is characterized by less variance than the other ENSO regions.  \\
    
     Figure 12 shows power spectrum corresponding to the intraseasonal daily Ni\~no 3 index for  1981-2018yy period (the data for the computations were taken from site \cite{nl}). The straight line is drawn to indicate correspondence to the Eq. (2) with $\beta =1/2$. The characteristic time scale obtained from the best fit is $T_0=1/f_0 \simeq 280d$. Figure 13 shows power spectrum corresponding to the intraseasonal daily Ni\~no 4 index for the 1981-2018yy period (the data for the computations were taken from site \cite{nl}). The straight line is drawn to indicate correspondence to the Eq. (2) with $\beta =1/2$. The characteristic time scale obtained from the best fit is $T_0=1/f_0 \simeq 196d$.  \\
   
   Comparing the Figs. 6, 7 with the Figs. 12, 13 one can conclude that the Hamiltonian distributed chaos observed in the ENSO is more similar to the chaos observed in the WNPM than to that observed in the ISM. This conclusion is consistent with the other, above mentioned, observations.  Moreover, compare the Fig. 10 with the Figs. 12 and 13 one can conclude that the Australian Monsoon belongs to an extended ENSO system (at least on the intraseasonal time scales).

\begin{figure}
\begin{center}
\includegraphics[width=8cm \vspace{-0.8cm}]{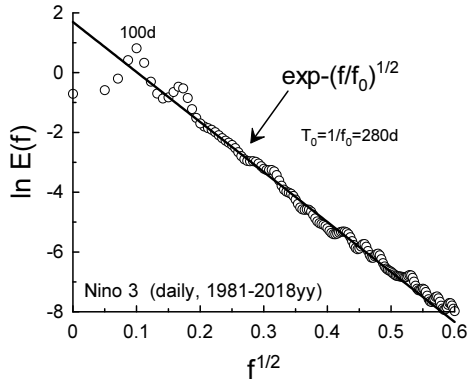}\vspace{-4cm}
\caption{\label{fig16} Power spectrum for the intraseasonal daily Ni\~no 3 index. } 
\end{center}
\end{figure}    

\begin{figure}
\begin{center}
\includegraphics[width=8cm \vspace{-0.77cm}]{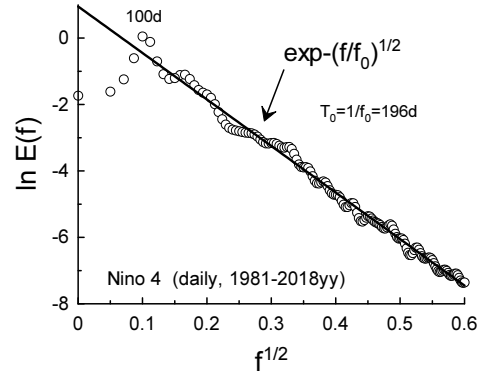}\vspace{-4.03cm}
\caption{\label{fig17} Power spectrum for the intraseasonal daily Ni\~no 4 index.  } 
\end{center}
\end{figure}

\section{Acknowledgement}

I thank H.J. Fernando and B. Galperin for inspiring comments and J.C. Sprott for sharing 
his data. I acknowledge use of the data provided by the KNMI Climate Explorer, by the Australian  Bureau of Meteorology, and by the NOAA: the Climate Prediction Center and  the Asia-Pacific Data Research Center at the University of Hawaii.


\begin{thebibliography}{99}
\bibitem{wwl} B. Wang, R. Wu and K.-M. Lau, J. Climate, {\bf 14}, 4073 (2001).
\bibitem{wf} B. Wang and Z. Fan,  Bull. Amer. Meteor. Soc., {\bf 80}, 629 (1999).
\bibitem{mon} http://apdrc.soest.hawaii.edu/projects/monsoon/
\bibitem{global} http://www.cpc.ncep.noaa.gov/products/Global\_Monsoons
\bibitem{b} A. Bershadskii, Phil. Trans. R. Soc. A, {\bf 371}, 20120168 (2013).
\bibitem{lin} H. Lin, J. Atmos. Sci. {\bf 66}, 2697 (2009).
\bibitem{ding} Q. Ding, Q et al., 2011. J. Climate, {\bf 24}, 1878 (2011).
\bibitem{gw} G. Grunseich and B. Wang, J. Climate, {\bf 29}, 9097 (2016).
\bibitem{fm}U. Frisch and R. Morf, Phys. Rev., {\bf 23}, 2673 (1981).
\bibitem{f} J.D. Farmer, Physica D, {\bf 4}, 366 (1982).
\bibitem{oh} N. Ohtomo, K. Tokiwano, Y. Tanaka et. al., J. Phy. Soc.
Jpn. {\bf 64} 1104 (1995).
\bibitem{sig} D.E. Sigeti, Phys. Rev. E, {\bf 52}, 2443 (1995).
\bibitem{b1} A. Bershadskii, arXiv:1803.10139 (2018).
\bibitem{suz} R.Z. Sagdeev, D.A. Usikov, G.M. Zaslavsky, Nonlinear Physics: from the 
Pendulum to Turbulence and Chaos (Harwood, New York, 1988).
\bibitem{b2} A. Bershadskii, arXiv:1801.07655 (2018).
\bibitem{spot} J.C. Sprott, Chaos and Time-Series Analysis (Oxford. University Press, 2003).
\bibitem{gen} http://sprott.physics.wisc.edu/cdg.htm
\bibitem{as} S. Ashkenazi and V. Steinberg, Phys. Rev. Lett. {\bf 83}, 3641 (1999).
\bibitem{tg} C. Tong and A. Gluhovsky, Phys. Rev. E {\bf 65}, 046306 (2002).
\bibitem{gl} A. Gluhovsky, Nonlinear Processes in Geophysics, {\bf 13}, 125 (2006).
\bibitem{she} T.G. Shepherd, Advances in Geophysics, {\bf 32},  287 (1990).
\bibitem{sh} T.G.Shepherd, Encyclopedia of Atmospheric Sciences, J. R. Holton et al., Eds.,
929 (Academic Press, 2003).
\bibitem{pel} V. Pelino et al., Commun. Nonlinear Sci. Numer. Simulat., {\bf 17}, 2122 (2012).
\bibitem{gl2} A. Gluhovsky, and K. Grady, Chaos, {\bf 26}, 023119 (2016).
\bibitem{b3} A. Bershadskii, arXiv:1804.08536 (2018).
\bibitem{ven2} M.J. Ventrice et al., Monthly Weather Review, {\bf 141}, 4197 (2013).
\bibitem{kim} K.Y. Kim et al., J. Geophys. Res., {\bf 111}, D20105 (2006).
\bibitem{sup} R. Suppiah, International Journal of Climatology, {\bf 24}, 269 (2004).
\bibitem{cli} http://www.clivar.org/asian-australian-monsoon
\bibitem{ka} Y. Kajikawa, B. Wang and J. Yang, International Journal of Climatology, {\bf 30}, 1114 (2010).
\bibitem{wang2} B. Wang, I.S. Kang and J.Y. Lee, Journal of Climate,
{\bf 17}, 803 (2004).
\bibitem{reg} http://www.cpc.ncep.noaa.gov/products/analysis\_monito
ring/ensostuff/nino\_regions.shtml
\bibitem{kg} V. Krishnamurthy and B. N. Goswami, J. Climate, {\bf 13}, 579 (2000).
\bibitem{wang} B. Wang et al., Geophys. Res. Lett., {\bf 32}, L15711 (2005).
\bibitem{chou} C. Chou, J.Y. Tu and J.Y. Yu, J. Climate, {\bf 16}, 2275 (2003),
\bibitem{jiang} W. Jiang et al., J. Climate, {\bf 30}, 109 (2017).
\bibitem{gre} J. Cr\'etat et al., Climate Dynamics, {\bf 49}, 1429 (2017).
\bibitem{nl} https://climexp.knmi.nl/start.cgi


\end{thebibliography}
\end{document}